\documentclass[11pt]{article}

\usepackage{lineno}
\usepackage{amsmath}
\usepackage{graphicx}
\newlength{\figwidth}
\begin{document}

\bibliographystyle{RS}

\title{Differential cellular contractility as a mechanism for stiffness sensing}
\author{Carina M Dunlop\\
Centre for Mathematical and Computational Biology,\\ Department of Mathematics,\\
University of Surrey, Guildford, Surrey, GU2 7XH\\
Email: c.dunlop@surrey.ac.uk}
\date{}
\maketitle
\begin{abstract}
The ability of cells to sense and respond to the mechanical properties of their environments is fundamental to a range of cellular behaviours, with substrate stiffness increasingly being found to be a key signalling factor. Although active contractility of the cytoskeleton is clearly necessary for stiffness sensing in cells, the physical mechanisms connecting contractility with mechanosensing and molecular conformational change are not well understood. Here we present a contractility-driven mechanism for linking changes in substrate stiffness with internal conformational changes. Cellular contractility is often assumed to imply an associated compressive strain. We show, however, that where the contractility is non-uniform, localized areas of internal stretch can be generated as stiffer substrates are encountered. This suggests a physical mechanism for the stretch-activation of mechanotransductive molecules on stiffer substrates. Importantly, the areas of internal stretch are not co-localized with those traditionally associated with stiffness sensing, e.g. focal adhesions, supporting recent experimental results on whole-cell mechanically-driven mechanotransduction. Considering cellular shape we show that aspect ratio acts as an additional control parameter, so that the onset of positive strain moves to higher stiffness values in elliptical cells.\end{abstract}

\section{Introduction}
It is clear that the mechanical properties of cell environments play a crucial role in controlling and coordinating cell behaviours both individually and within multicellular tissues. It has been specifically observed that substrate stiffness has a significant influence on phenotype across a range of cell types, for example, experiments have shown that stem cells can alter their differentiation target \cite{Engler}, cardiomyocytes de-differentiate and initiate proliferation \cite{GeigerELife}, and fibroblasts change their DNA synthesis and undergo apoptosis \cite{DemboFibro} in response to changes in the stiffness they encounter.  Active contractility of the cell cytoskeleton is the key mechanism by which cells physically interact with their environments and actomyosin contractility is found to underpin mechanotransduction across studies \cite{Wozniak,Jain,DemboFibro}. Consequently, a range of experimental techniques have been developed to measure cell-derived forces including traction force microscopy and other similar deformation-based approaches \cite{Polacheck,Style}. These, in combination, with substrates of carefully calibrated stiffness enable quantitative investigations into contractility and stiffness sensing \cite{Trichet,Liu}.

Despite this awareness of the necessity of cellular contractility for mechanotransduction the details of the coupling of the physical and molecular mechanisms are still unclear with a range of pathways and networks implicated \cite{Engler,Murrell,SchwarzGardell} with much current interest in the Hippo network and the associated YAP/TAZ transcriptional regulators \cite{Low,Yu}. Structurally, most experimental efforts have focused on the focal adhesions (FAs) as potential sites of force transduction. Focal adhesions are localised patches of strong cellular adhesion that are observed to form through integrin recruitment on comparatively stiff gel substrates \cite{MolecularBiology}. The potential of FAs as force transducers through stretch-activation and downstream signalling has been clearly demonstrated \cite{GeigerReview}. Despite this there is an new and emerging realisation that mechanotransduction is a whole cell process and that mechanically induced intracellular signalling cannot be explained by downstream biochemical signalling from stretch-activated FAs alone \cite{PNASWang,Twisting,NucleusReview, NucleusReview2}. The focus is on identifying potential mechanical and physical mechanisms for coupling the ECM directly to intracellular force sensing elements. Na \emph{et. al} \cite {PNASWang} have demonstrated that the speed of activation of Src molecules and the colocalisation of this activation with microtubule deformation implies a physical mechanism for transmitting force to internal stretch-activation molecules. There has also been recent direct evidence of stretch-activation in chromatin \cite{Twisting,NucleusReview2} and at the nuclear envelope \cite{NucleusReview, NucleusReview3}. 

There are several theoretical approaches currently used for describing cell mechanics, although these may be broadly separated into computational simulations and active matter representations. Computational simulations of cellular forces tend to focus on the dynamics of subcellular components looking at, for example, how these coordinate to control shape, cytoskeletal organisation and force generation, see e.g. \cite{JTBFly,McGarry,Weichsel,Albert,Afines}. Such subcellular focused simulations have also been used to model adherent cells with a focus on stiffness mediated effects\cite{Borau1,Cao,Pathak,Milan}. Active matter type representations in contrast use continuum modelling to describe the cell, with subcellular processes incorporated through phenomenological terms in the model \cite{Joanny}. In this context, cellular contractility can be modelled by e.g. an analogy with thermoelasticity and thermal cooling \cite{PRL,Joanny,MarchettiModel}. This approach that has been successfully used across a range of applications \cite{PRL,MarchettiModel,Oakes,Vichare,Mullen,He,Zemel}, including to explain the stiffness mediated changes in measured tractions \cite{Marcq} and force polarisation \cite{Friedrich}.

We are interested here in the potential for cellular contractility to drive mechanotransduction in cells and adopt a thermoelasticity-based active matter framework to explore this. Motivated by the observation that actomyosin complexes are not uniformly distributed throughout the cell \cite{Murrell, SchwarzGardell} we focus on variations in contractility throughout the cell showing how this differential contractility could drive force transmission from the substrate to the internal cellular structures. We further show how, as the stiffness of the extracellular substate passes a critical threshold, this contractility-based mechanism will drive stretch activation within the cell. Indeed, in contrast to the common assumption that cellular contractility induces internal contraction where contractility increases towards the edge of the cell a qualitative change can occur. Whereas on sufficiently soft substrates the strains are always negative as expected, in contrast on stiffer substrates this can switch so that internal stretch is generated within the cell. This suggests a physical mechanism underpinning internal mechanotransduction as stretch-activated molecules will be activated on encountering these stiffer substrates. Importantly, the region of cytoskeletal tension that is generated by this mechanism is distanced from the areas associated with focal adhesions. We further explore the role of cellular anisotropy in tuning this internal response mechanism.

\section{Mathematical model of differential contractililty in a cell on a deformable substrate}
We assume that the cell is spread and that it adheres to the underlying substrate; this is usually ensured experimentally by coating the surface with e.g. fibronectin or collagen \cite{Style,Polacheck}. Thus adopting a plane stress approximation and making the thermoelastic analogy for cellular contractility \cite{PRL} the internal cellular stresses $\mathbf{F}$ within the cell and the cellular strains (stretch) $e_{ij}$ can be related by
\begin{equation}
\label{eq:Constitutive}
F_{ij} = \frac{h E_\mathrm{c}}{1+\nu}\left( e_{ij}+\frac{\nu}{1-\nu}e_{kk}\delta_{ij} \right)
-\frac{h E_\mathrm{c}}{2(1-\nu)} P \delta_{ij},
\end{equation}
where is $E_\mathrm{c}$ the cellular Young's modulus and $\nu$ the cellular Poisson's ratio. The stress and strain is averaged over the cell height $h$ in plane stress so that these tensors are two dimensional. The constants in the second term are chosen so that in the absence of stress $e_{ii}=P$  (summation convention applies), and we see that $P$ represents the target area change of a material element as required. The internal cellular stresses are determined from the force balance 
\begin{eqnarray}
\nabla\cdot \mathbf{F} -KN \mathbf{u}&=&\mathbf{0}, 
\label{eq:Equilibrium}
\end{eqnarray}
where the underlying substrate is modelled as an array of linear springs, with $\mathbf{u}$ the displacements and $K$ the effective spring constant with $N$ the number density of springs. This approximation for the material response of the substrate is widely adopted for such contractility models with much success, see e.g. \cite{MarchettiModel, PRL}, and can be formally justified for thin gel layers \cite{ThinGel} or as an exact representation of the widely-used microstructured pillar arrays \cite{TanChen, Polacheck}.  If the cell boundary is denoted $\partial D$ the boundary condition
\begin{equation}
\mathbf{F} \cdot \mathbf{n} = \mathbf{0} \ \textrm{on} \ \partial D
\label{eq:zerostress}
\end{equation}
is applied to ensuring zero applied stress, whereas integrating Eq.~(\ref{eq:Equilibrium}) over the cell area $D$
and applying the zero-stress boundary condition also gives 
\begin{equation*}
\int_{D} K N \mathbf{u}=\mathbf{0}.
\end{equation*}

\subsection{Simplification of model for cells on soft substrates exhibiting circular geometry.}  On soft substrates cells are often quasi-circular and with a significant increase in actin density nearer the edges of the cell clearly observable \cite{GeigerReview}. Such circular cells shapes and cytoskeletal organisation can also be observed on stiffer substrates where the substrate is patterned in some way to compel the cell to take up a particular geometry \cite{Oakes, Murrell}.

Adopting this geometry, we take the cell to be a circular disc of radius $r_0$ and take  $P=P(r)$, where $r$ is the distance from the cell centre, and assume that $dP/dr\leq0$ and $P(r)\leq0$ for all $r$ so that the cell is uniformly contractile with increasing contractility towards the cell edge. Symmetry implies that the displacement $\mathbf{u}=u(r)\mathbf{e}_r$, i.e. all displacements are in the radial direction. Substituting (\ref{eq:Constitutive}) into (\ref{eq:Equilibrium}) in this case then gives
\begin{eqnarray}
r^2 \frac{d^2 u}{d r^2} +r\frac{d u}{d r}-\left(1+\gamma^2 r^2 \right) u&=&\frac{1}{2}\left(1+\nu \right) r^2 \frac{d P}{d  r},
\label{eq:radial}
\end{eqnarray}
with boundary conditions
\begin{equation}
u=0 \ \ \textrm{at } r=0, \ \ \frac{d u}{d r}+\nu \frac{u}{r}=(1+\nu)\frac{P}{2} \ \ \textrm{at } r=1.
\label{eq:radialbc}
\end{equation}
In (\ref{eq:radial}) and (\ref{eq:radialbc}) we have non-dimensionalised with the cell radius so that $u= r_0 \hat{u}$ and $r= r_0 \hat{r}$ and have subsequently dropped the hats. The non-dimensional parameter $\gamma^2=r_0^2 K N (1-\nu^2)/hE_c$ captures the relative elastic responses of the substrate and the cell. 

We note that (\ref{eq:radial}) permits an exact solution as the homogenous equation is solved by modified Bessel functions \cite{Abramowitz} so that by the method of variation of parameters we find that
\begin{multline}
u= A I_1(\gamma r) +B K_1(\gamma r) \\ +  \frac{2}{1+\nu} \int_0^{r} I_1(\gamma r) K_1(\gamma t) t\frac{d P}{d t} d t  \\+ \frac{2}{1+\nu} \int_{r}^1 K_1(\gamma r) I_1(\gamma t) t\frac{d P}{d t} d t
\label{eq:analyticalSolution}
\end{multline}
where the constants $A$ and $B$ are determined from the boundary conditions. However, due to the complexity of these expressions for our initial analysis it is more convenient to numerically solve the boundary value problem (\ref{eq:radial}) for particular $P$; all numerical solutions are obtained using the spectral-based MATLAB suite chebfun (v5.5.0) \cite{chebfun}. 

\section{Results}
\subsection{Differential contractility can generate internal stretch away from areas of maximum force transmission}
\begin{figure}
\centering{\includegraphics[width=0.7 \textwidth]{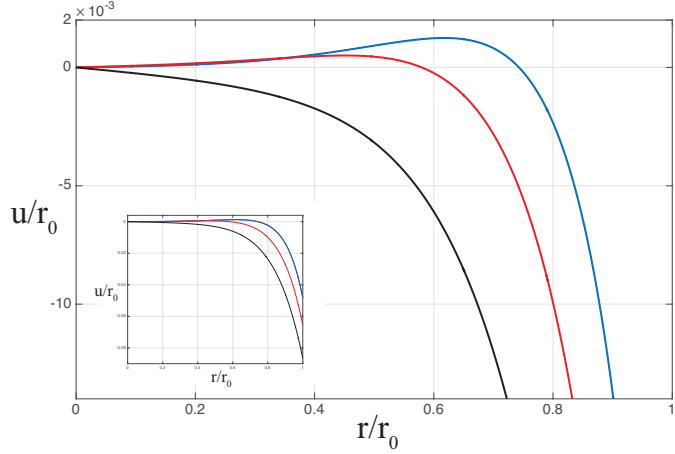}}
\caption{Plot of displacement $u$ for a cell with radius $r_0$ and contractility profile $P=-0.4(1+r^5)$ with $\gamma=5,7,10$ from bottom to top with $\nu=0.45$. Inset shows displacement profile over the entire disc with $\gamma=5,7,10$ from bottom to top as before.}
\label{fig:FIG2}
\end{figure}
We begin by taking a contractility profile of the form $P=-a(1+b r^5)$ to model the distribution of contractility throughout the cytoskeleton. The solution of (\ref{eq:radial}) in this case is plotted in Fig.~\ref{fig:FIG2} for $\gamma=5,7,10$ with $a=0.4$  and $b=1$. With $\gamma=5$, we see that all displacements are inwards, and additionally, the cell can be seen to be uniformly under negative strain as the gradient of the curve is uniformly negative. However, as the relative stiffness of substrate that the cell encounters increases (i.e. $\gamma$ increases) we see that the displacement profile goes through a qualitative transition. For these stiffer substrates, the increased contractility at the edge pulls the interior outwards towards it, resulting in positive displacements. In tandem with these positive displacements, there now exist regions within the cell that experience positive strains which puts the cytoskeleton under stretch with the consequent potential for stretch-activated mechanotransduction. 

Considering, for example, the stretch in the radial direction as given by the strain $e_{rr}=\partial u/\partial r$ with $\gamma=10$ we see from Fig.~\ref{fig:FIG2} that the strain increases from the origin to a maximum beyond which it decreases to zero strain before becoming negative. Interestingly the point of maximum strain is significantly set back from the cell periphery so that the region of stretching is not co-localised with the maximum mechanical activity and application of traction forces at the cell edges. Note that this stretch would be felt almost instantaneously as the stiffer substrate was encountered as it is a mechanical result of the equilibrium force balance within the cell. In this case that the mechanical coupling inherent in this model captures the transmission of stiffness information instantaneously to the interior of the cell as observed in \cite{PNASWang,Twisting,NucleusReview, NucleusReview2}, rather than there being a delay as would be expected in the case of a biochemical signal being transmitted from the sites of integrin adhesion.
\begin{figure}
\centering{\includegraphics[width=0.45\textwidth]{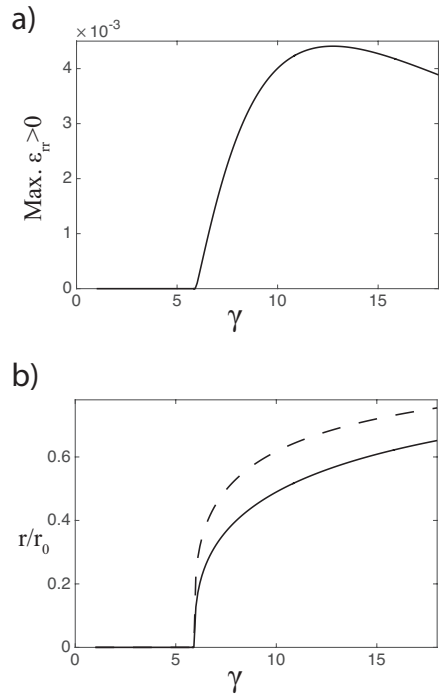}}
\caption{a) Plot of maximum intracellular radial stretch against $\gamma$; where the cell is uniformly in experiencing negative strain radially this is set to zero. Increasing $\gamma$ implies increased substrate stiffness. b) Plot of radius at which maximum radial strain achieved (solid line) and radius at which zero radial strain experienced (dashed line) against $\gamma$. The edge of the cell is at $r/r_0=1$. In both a) and b) $P=-0.4(1+r^5)$ and $\nu=0.45$.}
\label{fig:FIG3}
\end{figure}

Exploring further how the substrate elasticity affects the within-cell strain we plot in Fig.~\ref{fig:FIG3}(a) the maximum positive radial strain (where the strain is purely negative the value is set to zero). It is clear that for softer substrates (small $\gamma$) all strains are negative so that the cell is universally contracting. However, as the substrate gets stiffer and $\gamma$ increases we reach a critical point $\gamma=\gamma_c\approx5.9$ where there are now regions of the cell that are experiencing stretch. The maximum strain depends non-linearly on $\gamma$ for although as the relative stiffness of the substrate increases it enables positive strains it also increases the substrate resistance to deformations. The radius at which maximum radial tension is achieved is plotted in Fig.~\ref{fig:FIG3}(b) (solid line) where it is seen that as $\gamma$ increases so this region of stretch moves towards the outer edge of the cell. Additionally, we plot in Fig.~\ref{fig:FIG3}(b) the radius at which zero strain is achieved (dashed line) which marks the turning point as the strain returns to compression at the cell edge.

\subsection{Altering the contractility distribution tunes the strain-switching point of individual cells}
\begin{figure}
\centering{\includegraphics[width=0.7\textwidth]{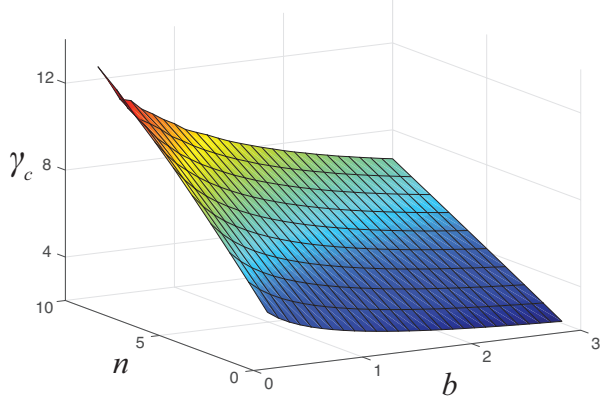}}
\caption{Plot of the critical value $\gamma_c$ at which positive radial strains are generated within cells for contractility profiles $P=-(1+br^n)$ with $\nu=0.45$.}
\label{fig:FIG4}
\end{figure}

By altering the internal distribution of contractility, cells would be able to tune the value of $\gamma$ at which a stretch-based mechanotransductive switching occurs. To show this we continue to consider contractility profiles of the general form $P=-a(1+br^n)$ and first note that $a$ can be scaled out of the system by letting $u=a\tilde{u}$, which leaves the point at which strain switching occurs $\gamma_c$ unaltered. The parameter $a$ does control the magnitude of the displacements and strains experienced and so is a key fit parameter when comparing with experimental data. However, in analysing the critical switching point of the system where the strain reaches zero we without loss of generality set $a=1$ in the analysis. Plotting the critical value of $\gamma_c$ against $n,b$ in Fig.~\ref{fig:FIG4} we see that as $n$ and $b$ increase and decrease, respectively, so the value of $\gamma_c$ increases. Thus by, for example, concentrating contractile activity nearer the edge of the cell (i.e. $n$ increasing) the stiffness at which switching occurs can be increased. It can be shown that, as expected, where positive displacements are generated within the cell these will be in a region $r \leq r_c$ for some $r_c<r_0$ with negative strains in $r_c<r\leq r_0$. Indeed this is true for all contractility profiles that monotonically increase in magnitude towards the cell edge, see Appendix A.

\subsection{Increasing cellular aspect ratio shifts switching point of differential contractility-based mechanotransduction}

We now proceed to consider the effect of increasing aspect ratio on this differential contractility-based mechanotransductive mechanism. It has been clearly demonstrated in a number of contexts that increasing the aspect ratio of a cell can increase cellular traction and additionally lead to increased internal stress, see e.g. \cite{BanerjeeNJP}, which raises the question of how such a change of geometry affects the mechanotransductive mechanism elucidated here. 

As the adhered area of the circular cells so far considered is $\pi r_0^2$ we fix the area of the elongated cells to this value and consider an elliptical cell of semi-major axis $\alpha r_0$ and semi-minor axis $r_0/\alpha$. Such elliptical cells are commonly generated in experiments using adhesive micropatterns to control cell geometry \cite{Oakes, Murrell}. The degree of shape anisotropy of the cell is controlled by $\alpha$ from which the eccentricity of the ellipse can be calculated as $\sqrt{1-\frac{1}{\alpha^2}}$. We thus solve (\ref{eq:Constitutive}) and (\ref{eq:Equilibrium}) on this geometry with the zero stress boundary conditions (\ref{eq:zerostress}) and $\mathbf{u}=0$ at the origin (by symmetry). The non-dimensional parameter $\gamma$ again quantifies the stiffness of the underlying substrate.  

\begin{figure}
\centering{\includegraphics[width=0.85\textwidth]{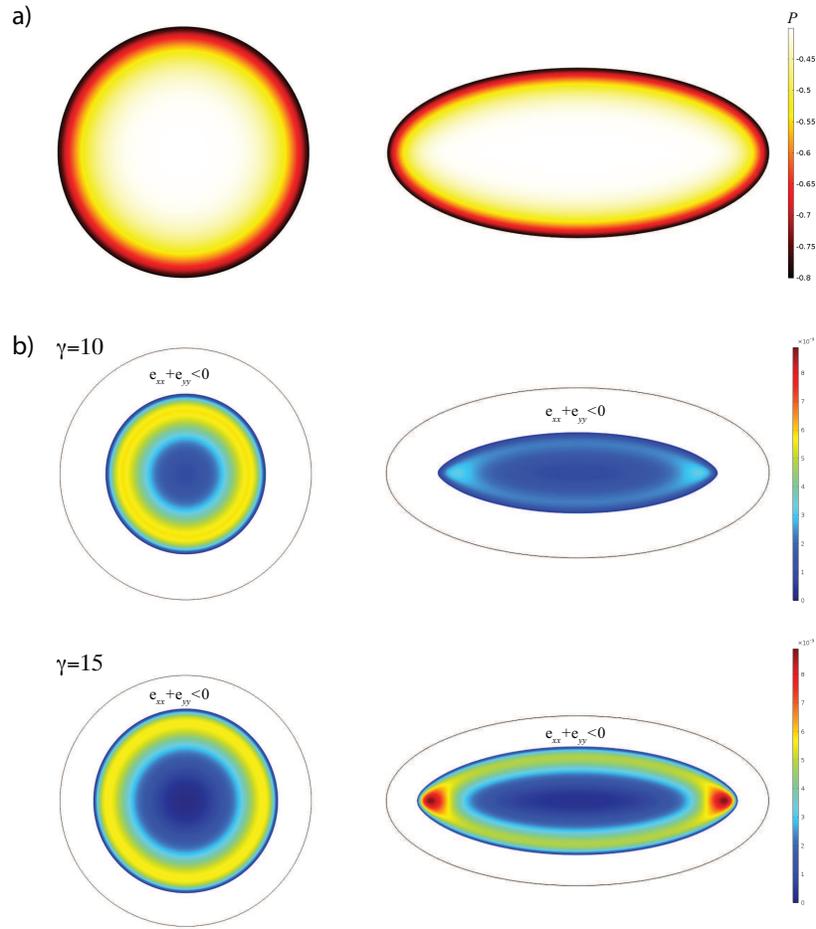}}
\caption{a) Heatmap of the contractility profile $P$ in a circular cell with $\alpha=1$ and an elliptical cell with $\alpha=1.5$. The greatest contraction occurs at the cell edge where $P=0.8$. b) Heatmap of the strain invariant $e_{xx}+e_{yy}$ in a circular cell with $\alpha=1$ and an elliptical cell with $\alpha=1.5$. The edge of the cell is indicated by a black line and regions without colour have $e_{xx}+e_{yy}<0$. }
\label{fig:fig5}
\end{figure}

The cellular contractility term $P$ in (\ref{eq:Constitutive}) is again the target area change of a small material element within the cell, however, now we assume that $P=P(d)$ where $d$ is the normal distance from the cell edge. We solve this system numerically using Finite Element Methods, COMSOL (COMSOL Multiphysics® v.5.2, Stockholm, Sweden), specifying the contractility in a way that is consistent with the approach taken to circular geometries. As such we take $P=-a(1+\left(1-\frac{\alpha d}{r_0} \right)^5)$ and plot in Fig.~\ref{fig:fig5}(a) a heatmap of the contractility profiles with $\alpha=1$ and $\alpha=1.5$. As before the maximum magnitude of contractility $a$ can be scaled out of the system and represents an experimental fit parameter determining the magnitude of strain. We here take a=0.4 throughout.

We choose to quantify strain in this fully two-dimensional model using $e_{ii}$ (an invariant of the strain tensor), which in component form may be expressed as $e_{xx}+e_{yy}$. This measure quantifies the amount a small area element is stretched, so that if negative the element's area decreases and if positive it is stretched. In Fig.~\ref{fig:fig5}(b) we plot a heat map of the regions where this measure of strain is positive for two different substrate stiffnesses both above the critical threshold for positive strains to exist. We consider $\gamma=10$ and $\gamma=15$. The results for the circular cell with $\alpha=1$ are consistent with those already presented with the region encompassing positive strains increasing with increasing stiffness and the radius of maximal strain moving out towards the edge of the cell as stiffness increases. In the elliptical cell, the effect of stiffness is broadly similar, however with some key differences. The anisotropic shape now enables hotspots of high strain to form as the stiffness increases, although on softer substrates the maximal strain is less than that experienced in a circular cell. The profiles of the displacements along the major and minor axes can be shown to follow profiles similar to those plotted in Fig.~\ref{fig:FIG2}, see Fig.~SI1.

\begin{figure}
\centering{\includegraphics[width=0.85\textwidth]{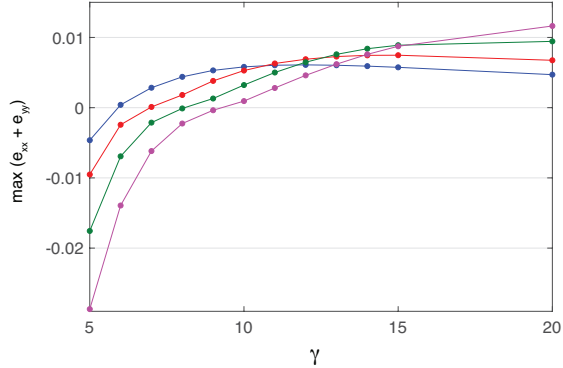}}
\caption{Plot of maximum total strain $e_{xx}+e_{yy}$ as a function of increasing substrate stiffness. The different curves represent cells of increasing anisotropy with $\alpha=1, 1.25,1.5,1.75$ in blue, red, green and purple, respectively (running top to bottom at $\gamma=5$).  }
\label{fig:fig6}
\end{figure}
We plot in Fig.~\ref{fig:fig6}, the maximum of $e_{xx}+e_{yy}$ across the cell as $\gamma$ increases for cells of increasing anisotropy. It can be seen that increasing the anisotropy delays the point at which positive strains are generated within the cell, so that elliptical cells can be expected to have a $\gamma_c$ higher than that for circular cells. This effect is monotonic increasing with $\alpha$. However, as substrate stiffness further increases so the maximum strain for an elliptical cell eventually overtakes that of a circular cell with, as can be seen from the heatmap, a focusing of strain into two hotspots on the long axis.

\section{Conclusions}
The ability of cells to respond to changes in the stiffness of their surroundings plays a crucial role in determining cell function. While focal adhesions (FAs) are undoubtably a crucial part of the mechanotransductive apparatus it has also become clear that there must be additional regions of physical force transduction within cells. Here we propose differential contractility as a mechanism for generating this internal mechanotransduction. Using a continuum elasticity approach we have shown that where contractility increases towards the cell edge this will generate positive strains and consequent cytoskeletal stretch on sufficiently stiff substrates. Importantly, the region of cytoskeletal stretch is not co-localised with FA situation or areas of maximum mechanical activity being instead set back from the cell periphery, sited much deeper within the cell. These results support the emerging realisation of the need to look for internal mechanotransductive stretch-activation across the whole cell. We have also shown that the critical stiffness at which internal stretch will be generated can be controlled by adjusting the contractility profile and is shifted by the introduction of anisotropy in the cell geometry. This transition to internal stretch is a purely mechanical consequence of the elastic coupling between the cell and the substrate and as such would enable an almost instantaneous response as the cell comes into mechanical equilibrium. The emergence of regions of stretch within the cell as stiffer substrates are encountered thus provides a potential physical mechanism for coupling the extracellular environment to internal molecular conformational change.

\section*{ACKNOWLEDGMENTS}

CMD acknowledges financial support from the UK Engineering and Physical Sciences Research Council (EP/M012964/1). The data underlying the findings are available, details of the data and how to request access are available from the University of Surrey.

\appendix
\section{For circular cells positive displacements are restricted to the central region}
To show that all positive displacements are restricted to the central region of the cell we return to the analytical solution (\ref{eq:analyticalSolution}), applying the boundary condition that $u\to0$ as $r\to 0$ gives that
\[B=-\frac{2}{1+\nu}\int_0^1 I_1(\gamma t)t \frac{d P}{d t} d t, \] 
which guarantees the correct limiting behaviour. Now we may express the solution (\ref{eq:analyticalSolution}) in the form 
\begin{multline}
u= I_1(\gamma r) \left(A +\frac{2}{1+\nu} \int_0^{r} K_1(\gamma t) t\frac{d P}{d t} d t \right)\\ -\frac{2}{1+\nu} K_1(\gamma r)\int_{0}^{r}  I_1(\gamma t) t\frac{d P}{d t} d t.
\label{eq:Interim2}
\end{multline}
we now follow \cite{Relton} and show that a solution of this form can only have one positive zero. We divide (\ref{eq:Interim2}) by $I_1(\gamma r)$, and differentiate with respect to $r$ to obtain (after recalling that $I_1(x)K_1(x)'-K_1(x)I_1(x)'=-1/x$ - see e.g. \cite{Abramowitz}) 
\begin{eqnarray}
\frac{d}{d r}\left( \frac{u(r)}{I_1(\gamma r)} \right ) &=& \frac{2}{I_1(\gamma r)^2r(1+\nu)}\int_0^{r} I_1(\gamma t)t \frac{d P}{d t} d t .
\label{eq:Analytical2}
\end{eqnarray}
For a monotonically decreasing contractility function with $dP/dr<0$ we see that the right hand side of (\ref{eq:Analytical2}) is always strictly less than zero. To complete the proof we choose an arbitrary interval $\lbrack \alpha,\beta \rbrack  \in ( 0,1\rbrack  $ and integrate  (\ref{eq:Analytical2}) over this region to obtain that
\begin{equation}
u(\beta)I_1(\gamma \alpha)-u(\alpha)I_1(\gamma \beta)<0.\label{eq:ReltonUnique}\end{equation}
The expression (\ref{eq:ReltonUnique}) can then be used to prove the uniqueness of any positive zero of the solution $u$. Assume there exist two such zeros such that $u(r_0)=0=u(r_1)$, $r_0< r_1$ then by taking $\alpha=r_0$ and $\beta=r_1$ we obtain a contradiction. As $u(1)<0$ (which can also be seen from (\ref{eq:ReltonUnique})) and there is only one zero in the region the result follows.

\end{document}